\documentstyle[sprocl,epsbox]{article}

\def\be{\begin{equation}}
\def\ee{\end{equation}}
\def\bea{\begin{eqnarray}}
\def\eea{\end{eqnarray}}

\newcommand\la{\langle}
\newcommand\ra{\rangle}

\def\slash#1{\rlap/{#1}}

\def\Sslash{\slash{\mkern-1mu S}}

\begin{document}
\title{RECENT DEVELOPMENTS IN PERTURBATIVE QCD:\\$Q^2$ EVOLUTION OF
CHIRAL-ODD DISTRIBUTIONS\\ $h_1(x,Q^2)$ and $h_L(x,Q^2)$
\footnote{Invited talk presented at ``QCD Corrections and the New Physics'',
October 27-29, 1997, Hiroshima, Japan.  To be published in the
proceedings. (ed. by J. Kodaira et al.)}
}
\author{YUJI KOIKE}
\address{Graduate School of Science and Technology, Niigata University\\
Ikarashi, Niigata 950-21, Japan\\E-mail: koike@nt.sc.niigata-u.ac.jp}
\maketitle
\abstracts{
After reviewing QCD definitions of the chiral-odd spin-dependent
parton distributions $h_1(x,Q^2)$ and $h_L(x,Q^2)$, I will summarize
the main feature of the recent two results in perturbative QCD:
(i) Next-to-leading order $Q^2$ evolution of 
$h_1(x,Q^2)$. (ii) Leading order
$Q^2$ evolution of the twist-3 distribution $h_L(x,Q^2)$ and the
universal simplification of 
the $Q^2$ evolution of all the twist-3 distributions
in the large $N_c$ limit.
}

\section{Introduction}

Recent developments in collider experiments
have been providing us with rich information on
the quark-gluon distributions in the nucleon.  
Particularly interesting are the spin structure functions
which reveal 
``spin distributions'' carried by
quarks and gluons inside the nucleon.
Besides the importance in their own right, they play an
indispensable role to test 
the spin-dependent part of QCD -- fundamental theory of
the strong interaction:  Deeper test
of QCD in the spin dependent level may eventually lead us
to search for a new physics beyond the standard model.

The nucleon's structure functions measured in hard processes
can be written as a sum of parton distributions for each
quark or anti-quark flavor and a gluon.
They are 
functions of Bjorken's $x$ which represent parton's momentum
fraction in the nucleon and a scale $Q^2$ at which they are measured.
Untill now, most data on the nucleon's structure functions
have been obtained through the lepton-nucleon 
deep inelastic scattering (DIS).  The chiral-odd
distributions,
$h_{1,L}(x,Q^2)$, however, can not be measured by the inclusive DIS,
and hence there has been no data up to now.
They can be measured by the nucleon-nucleon polarized Drell-Yan
process
and semi-inclusive DIS which detect particular hadrons
in the final state.   They will hopefully be measured by
planned experiments using  
polarized accelerators at BNL, DESY, CERN and SLAC etc\,\cite{SPIN}.  
In particular,
RHIC at BNL is expected to provide first data on these distributions.

In the study of these structure functions,
perturbative QCD plays an indispensable role in 
predicting their $Q^2$-dependence:  Given a structure function, 
say $h_1(x,Q_0^2)$, at one scale $Q_0^2$, perturbative QCD
predicts the shape of $h_1(x,Q^2)$ at an arbitrary scale $Q^2$.
This $Q^2$ evolution is necessary not only in extracting
low energy hadron properties from high energy experimental data
but also in testing 
the $x$-dependence 
predicted by a non-perturbative QCD technique or a model
with the high energy data.  
In this talk, I will first give the QCD definition
of $h_{1,L}(x,Q^2)$ and discuss their $Q^2$-dependence
studied in the recent literature.

\begin{figure}[h]
\epsfile{file=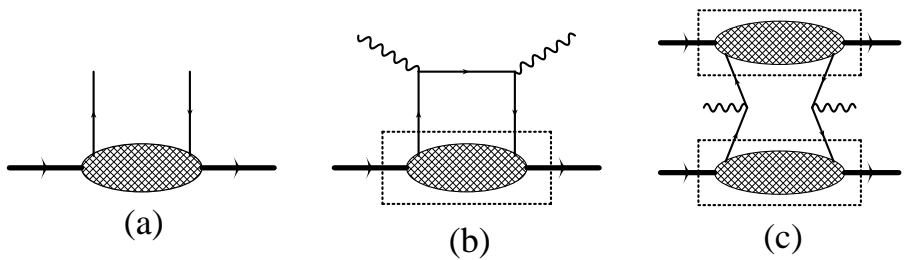,scale=0.63}
\caption[Fig. 1]{(a) Quark distribution function. 
(b) Nucleon struction function in DIS.  (c) Cross section for
the nucleon-nucleon Drell-Yan process.}
\end{figure}

\section{Chiral-Odd Distributions $h_{1,L}(x, Q^2)$}

Inclusive hard processes
can be generally analyzed in the framework 
of the QCD factorization theorem\,\cite{Mueller}.
This theorem generalizes the idea 
of the Bjorken-Feynman's ``parton model'' 
and allows us to
include QCD correction in a systematic way.
Here I restrict myself to the hard processes
with the nucleon target, 
such as lepton-nucleon deep inelastic
scattering (DIS, $l + p\to l' + X$), Drell-Yan ($p+ p'\to l^+l^- + X$),
semi-inclusive DIS ($l + p\to l' +h+ X$).
According to the above theorem, the cross section (or 
the nucleon structure function)
for these processes can be factorized into
a ``soft part'' and a ``hard part'':
The soft part represents the parton (quark or gluon) distribution
in the nucleon and the hard part describes
the short distance cross section between the parton 
and the external hard probe which is calculable within 
perturbation theory.
For example, 
a nucleon structure function in DIS
can be written as the imaginary part of the virtual photon-nucleon forward
Compton scattering amplitude. (Fig. 1 (b))
According to the above theorem, 
in the Bjorken limit, i.e. $Q^2, \nu=P\cdot q\to
\infty$ with $x=Q^2/2\nu =$ finite, ($Q^2=-q^2$ is the
virtuality of the space-like photon, $P$ is the nucleon's four momentum), 
the structure function can be written as
\bea
W(x,Q^2) = \sum_a \int_x^1 {dy\over y}H^a({x\over y}, {Q^2\over \mu^2},
\alpha_s(\mu^2))\Phi^a(y,\mu^2),
\label{DIS}
\eea
where $\Phi^a$ represents a distribution of parton $a$ in the nucleon
and $H^a$ describes the short distance cross section of the 
parton $a$ with the virtual photon.  
$\mu^2$ is the factorization scale.
In Fig. 1(b), $\Phi^a$
is identified by the dotted line. (Fig.1 (a)).
Similarly to DIS, the cross section for the nucleon-nucleon Drell-Yan 
process
can also be written in a factorized form
at $s=(P_A + P_B)^2, Q^2 \to \infty$ with a fixed $Q^2/s$
($P_{A,B}$ are the momenta of the two nucleons, $Q$ is the momentum of the
virtual photon): 
\bea
d\sigma \sim \sum_{a,b}\int_{x_a}^1 dy_a \int_{x_b}^1 dy_b H^{ab}
\left({x_a\over y_a}, {x_b\over y_b}, 
Q^2; {Q^2\over\mu^2}, \alpha_s(\mu^2)\right)
\Phi^a(y_a,\mu^2)\Phi^b(y_b,\mu^2),
\label{DY}
\eea
where
the two parton distributions, $\Phi^a$ and $\Phi^b$, 
for the beam and the target
appear as was shown by dotted lines in Fig. 1(c).

As is seen from Figs. 1(b),(c), the parton distribution can be regarded
as a parton-nucleon forward scattering amplitude
shown in Fig. 1 (a) which appear in several different hard processes.
In particular,
the quark distribution in the nucleon moving in the $+\hat{\bf e}_3$
direction can be written as the light-cone Fourier transform 
of the quark correlation function in the nucleon:\,\cite{CS}
\bea
\Phi^a(x,\mu^2)= P^+\int_{-\infty}^\infty {dz^- \over 2\pi}
e^{ixP\cdot z}\la PS | \bar{\psi}^a(0)\Gamma \psi^a(z)|_\mu |PS \ra,
\label{distribution}
\eea
where $|PS\ra$ denotes the nucleon (mass $M$) state 
with momentum $P^\mu$ 
and spin $S^\mu$, and $\psi^a$ is the quark field with flavor $a$.
In (\ref{distribution}),
we have suppressed for simplicity the gauge link operator
which ensures the gauge invariance and $|_\mu$ indicates the
operator is renormalized at the scale $\mu^2$.
A four vector $a^\mu$ is decomposed into two light-cone components
$a^\pm={1\over \sqrt{2}}(a^0 \pm a^3)$ and the transverse component
$\vec{a}_\perp$.  In (\ref{distribution}), $z^+=0$, $\vec{z}_\perp =\vec{0}$,
and $z^2=0$.  $\Gamma$ generically represents $\gamma$-matrices,
$\Gamma=\gamma_\mu, \gamma_\mu\gamma_5, \sigma_{\mu\nu}, 1$.
$\Phi^a(x,\mu^2)$ measures the distribution
of the parton $a$ to carry the momentum $k^+ = xP^+$
in the nucleon, which is independent from particular hard proceeses.

If one puts $\Gamma=\gamma_\mu,\gamma_\mu\gamma_5$, the chirality
of $\bar{\psi}$ and $\psi$ becomes the same, namely it
defines the chiral-even distributions.  Likewise,
putting $\Gamma=\sigma_{\mu\nu}, 1$ defines the chiral-odd 
disributions.  For the case of the deep-inelastic scattering
(Fig. 1 (b)),
the quark line emanating from the target nucleon comes back
to the original nucleon after passing through the hard interactions.
Since the perturbative interaction in the standard model preserves
the chirality except a tiny quark mass effect, the chirality of the
two quark lines entering the nucleon in Fig. 1(b) is
the same.   Hence the DIS can probe only the chiral-even
quark distributions.  
On the other hand, in the Drell-Yan process (Fig. 1 (c)),
there is no correlation in chirality between two quark lines
entering each nucleon.  Therefore the Drell-Yan process probes both chiral
even and odd distributions.

The chiral-odd distributions $h_1^a(x,\mu^2)$, $h_L^a(x,\mu^2)$
in our interest are defined by putting $\Gamma=\sigma_{\mu\nu}i\gamma_5$
in (\ref{distribution}):\,\cite{JJ}
\bea
& &P^+\int{dz^- \over 2\pi}
e^{ixP\cdot z}\la PS | \bar{\psi}^a(0)\sigma_{\mu\nu}i\gamma_5
\psi^a(z)|_\mu|PS\ra \nonumber\\
& &\quad\qquad
=2[h_1^a(x,\mu^2)(S_{\perp\mu}p_\nu - S_{\perp\nu}p_\mu)/M\nonumber\\
& &\quad\qquad+h_L^a(x,\mu^2)M(p_\mu n_\nu - p_\nu n_\mu)(S\cdot n)
\nonumber\\
& &\quad\qquad+h_3^a(x,\mu^2)M(S_{\perp\mu}n_\nu - S_{\perp\nu}n_\mu)]
\label{h1hL}
\eea
where we introduced two light-like vectors
$p$, $n$ ($p^2 = n^2=0$) by the relation
$P^\mu=p^\mu+ {M^2 \over 2}n^\mu$, $p\cdot n=1$, $p^-=n^+=0$.
If we write
$P^+={\cal P}$,
$p={{\cal P}\over \sqrt{2}}(1,0,0,1)$,
$n={1\over \sqrt{2}{\cal P}}(1,0,0,-1)$.
${\cal P}$ is a parameter which specifies the Lorentz frame of 
the system: ${\cal P}\to \infty$ corresponds to
the infinite momentum frame, and ${\cal P}\to M/\sqrt{2}$
the rest frame of the nucleon.
$S_\perp^\mu$ is the 
transverse component of $S^\mu$ defined by
$S^\mu=(S\cdot n)p^\mu+(S\cdot p)n^\mu + S_\perp^\mu$.
One can show that $\Phi^a$ defined in (\ref{distribution})
has a support $-1<x<1$.
If one replaces the quark field $\psi$ in (\ref{distribution})
by its charge conjugation field
$C\bar{\psi}^T$, it defines the anti-quark distribution
$\bar{\Phi}^a$.
In particular 
$h_{1,L,3}^a(x,\mu^2)$ in
(\ref{h1hL})
are related to their anti-quark distribution by
$h_{1,L,3}^a(-x,\mu^2)=-\bar{h}_{1,L,3}^a(x,\mu^2)$.

$\Phi^a$
appears in a physical cross section in the form of the 
convolution with a short distance cross section in a parton level
as is shown in (\ref{DIS}) and (\ref{DY}).
The cross section can be expanded in powers of ${1 \over \sqrt{Q^2}}$ 
as
\bea
\sigma(Q^2)\sim A({\rm ln}Q^2) +
{M\over  \sqrt{Q^2}}B({\rm ln}Q^2) +
{M^2\over  {Q^2}}C({\rm ln}Q^2)+\cdots,
\label{twistex}
\eea
where each coefficient $A$, $B$, $C$ receives
logarithmic $Q^2$-dependence due to the QCD radiative correction.
In order to see how 
$h_{1,L,3}$ can contribute in the expansion (\ref{twistex}), it is
convenient to move into the infinite momentum frame
(${\cal P}\sim Q\to \infty$).
In this limit the
coefficient of 
$h_{1,L,3}$ in (\ref{h1hL}) behaves, respectively, as
$O(Q)$, $O(1)$, $O(1/Q)$.   Therefore if $h_1$
contributes to the $A$ term in (\ref{twistex}), 
$h_L$ can contribute at most to the $B$-term, and $h_3$ 
can contribute at most to the $C$-term.
In general, when a distribution function contributes to 
hard processes at most in the order of
$\left(1\over \sqrt{Q^2}\right)^{\tau-2}$,
the distribution is called twist-$\tau$.
Therefore 
$h_1$, $h_L$, $h_3$ in (\ref{h1hL}) is, respectively,
twist-2, -3 and -4.   

Twist-2 distribution $h_1$ can be measured through the
transversely polarized Drell-Yan\,\cite{RS,AM,JJ,CPR}, 
semi-inclusive deep inelastic scatterings
which detect pion\,\cite{JJ93}, polarized baryons\,\cite{AM,Ji94,J96},
correlated two pions\,\cite{JJT}.

From the discussion above, one sees that 
it is generally difficult to isolate experimentally  
higher twist ($\tau \geq 3$) distributions in hard proceeses,
since they are hidden by the leading twist-2 contribution ($A$ term in
(\ref{twistex})).  However, this is not the case for $h_L$ and $g_T$.
In particular spin asymmetries,
they contribute to the $B$-term in the absence of $A$-term: 
$g_T$ can be measured in
the transversely polarized DIS \cite{Jg2}, and
$h_L$ appears in the longitudinal versus transverse spin asymmetry
in the polarized nucleon-nucleon Drell-Yan process\,\cite{JJ}.
Therefore the $Q^2$-evolution of $g_T$ and $h_L$
can be a new test of perturbative QCD beyond the twist-2 level.

Insertion of other $\gamma$-matrices in
(\ref{distribution}) defines other distributions.
In Table 1, we show the classification of the
quark distributions up to twist-3.\cite{JJ}
There
$f_1$, $g_{1,T}$, $e$ is defined, respectively, by 
$\Gamma=\gamma_\mu, \gamma_\mu\gamma_5, 1$ in (\ref{distribution}).
A similar classification can also be extended to the
gluon distributions\,\cite{MJ}.
The distribution $f_1$ contributes to the spin averaged 
structure functions
$F_{1,2}(x,Q^2)$ familiar in DIS.
The helicity distribution $g_1$ contributes
to the $G_1(x,Q^2)$ 
structure function measured in the longitudinally polarized
DIS.
By now there has been much accumulation of
experimental data on 
$f_1$ and $g_1$, and 
the data on $g_1$ triggered lots of theoretical discussion on the 
``origin of the nucleon spin''\,\cite{SPIN}.
The first nonzero data on $g_2$ ($=g_T-g_1$) 
was also reported in Ref. \cite{g2}.

\begin{table}
\begin{center}
\begin{tabular}{|c|ccc|}
\hline
spin& average  & longitudinal  & transverse \\ \hline
twist-2 & $f_{1}$ & $g_{1}$ & \underline{$h_{1}$} \\
twist-3 & \underline{$e$} &  \underline{$h_{L}$} & $g_{T}$ \\ \hline
\end{tabular}
\end{center}
\caption{
Clasification of the quark distributions based on spin, twist and chirality.
Underlined distributions are chiral-odd. Others are chiral-even.
}
\label{tab:1}
\end{table}

\section{Next-to-leading order (NLO) $Q^2$-evolution of $h_1(x,Q^2)$}

As we saw in the previous section, $h_1$ is the third and 
the final twist-2 quark 
distribution.  It has a simple parton model interpretation
as can be seen by the Fourier expansion of $\psi$ in 
(\ref{h1hL}).
It measures the probability in the
transversely polarized nucleon to
find a quark polarized parallel to the nucleon spin minus
the probability to find it oppositely polarized.
Here the transverse polarization refers to
the eigenstate of the
transverse Pauli-Luba\'{n}ski operator $\gamma_5\Sslash_\perp$.
If one replaces the transverse polarization
by the longitudinal one, 
it becomes the helicity distribution $g_1$.
For nonrelativistic quarks, $h_1(x,\mu^2)=g_1(x,\mu^2)$.
A model
calculation suggests, $h_1$ is the same order as 
$g_1$.\,\cite{JJ,PP,Jaffe97}

The $Q^2$-evolution of $h_1$ is described by the usual DGLAP
evolution equation\cite{DGLAP}.  Because of its chiral-odd nature
it does not mix with gluon distributions.
Therefore the $Q^2$-dependence of $h_1$ is 
described by the same equation both for singlet and nonsinglet
distributions.
For $f_1$ and $g_1$, the NLO $Q^2$ evolution
was derived long time ago\,\cite{FRS,GLY,CFP,MN}
and has been frequently used
for the analysis of experiments\,\cite{GRV,GRV2}.
The leading order (LO) 
$Q^2$-evolution for $h_1$ has been known for some time\,\cite{AM}.
In the recent literature, the next-to-leading order (NLO)
$Q^2$-evolution has
been completed by two papers\,\cite{Vog,HKK} almost at the same time: 
Vogelsang\cite{Vog} presented the light-cone gauge calculation for the
two-loop splitting function of $h_1$ in the formalism
originally used for $f_1$\cite{CFP}.
We \cite{HKK} carried out the Feynman gauge calculation
of the two-loop anomalous dimension following the method 
of Ref. \cite{FRS}
for $f_1$.   
The results of these calculations in the $\overline{\rm MS}$ scheme 
agreed completely.  
This result was subsequently confirmed by Ref. \cite{KM}.
In the following, I briefly discuss
the characteristic feature of the NLO $Q^2$ evolution of $h_1$
following Refs.\,\cite{HKK,HKK2}.

Analysis of (\ref{h1hL}) gives the connection between
the $n$-th moment of $h_1$ and a tower of
twist-2 operators:
\bea
{\cal M}_n[h_1(\mu^2)]&\equiv & \int_{-1}^1dx\,x^n h_1(x,\mu^2) = 
{ -1 \over 2M}  
\la PS_\perp | O^\nu_n(\mu^2)S_{\perp\nu} | PS_\perp \ra,
\nonumber\\
O^{\nu}_n&=&
\bar{\psi}\sigma^{\nu\alpha}n_\alpha
i\gamma_5 (in\cdot D)^n\psi,
\label{tw2}
\eea
where $S_\perp$ stands for the transverse polarization 
and $O^\nu_n(\mu^2)$ indicates the operator
$O_n^\nu$ is renormalizaed at the scale $\mu^2$.
The contraction with $n^\mu$ 
and $S_{\perp}^\mu$ (recall $S_\perp\cdot n=0$,
$n^2=0$)
in (\ref{tw2})
projects out the relevant twist-2 contribution from the composite
operator.  (``Twist'' for local composite operators is defined
as dimension minus spin.)
By solving the renormalization group 
equation
for $O^\nu_n S_{\perp\nu}$, one gets  
the NLO $Q^2$ dependence of 
${\cal M}_n[h_1(\mu^2)]$ as
\bea
{ {\cal M}_n[h_1(Q^2)] \over {\cal M}_n[h_1(\mu^2)]}
=\left( {\alpha_s(Q^2) 
\over \alpha_s(\mu^2)} \right)^{\gamma_n^{(0)}/2\beta_0}
\left[ 1 + { \alpha_s(Q^2) - \alpha_s(\mu^2) \over 4\pi }
{\beta_1\over \beta_0} \left( {\gamma^{(1)}_n \over 2\beta_1}-
{\gamma^{(0)}_n \over 2\beta_0 } \right) \right],\nonumber\\
\label{eq6}
\eea
where $\alpha_s(Q^2)$ is the NLO QCD running coupling constant 
given by 
\bea
{\alpha_s(Q^2) \over 4\pi} = {1 \over \beta_0{\rm ln}(Q^2/\Lambda^2)}
\left[ 1 - { \beta_1 {\rm lnln}(Q^2/\Lambda^2) \over \beta_0
{\rm ln}(Q^2/\Lambda^2)} \right],
\label{eq7}
\eea
with the
one-loop and two-loop
coefficients of the $\beta$-function
$\beta_0=11-2/3N_f$ and $\beta_1=102-38/3N_f$ ($N_f$ is
the number of quark flavor) and the QCD scale parameter $\Lambda$.
$\gamma^{(0)}_n$ and $\gamma^{(1)}_n$ are the 
one-loop and two-loop coefficients of the anomalous dimension $\gamma_n$ for
$O^{\nu}_nS_{\perp\nu}$
defined as
\bea
\gamma_n={\alpha_s \over 4\pi} \gamma_n^{(0)}
+\left({\alpha_s \over 4\pi}\right)^2 \gamma_n^{(1)}+\cdots.
\label{eq8}
\eea
If one sets $\beta_1\to 0$ and $\gamma^{(1)}_n\to 0$ in (\ref{eq6}),
the leading order (LO) $Q^2$ evolution is obtained.
$\gamma_n^{(0)}$ and
$\gamma_n^{(1)}$ are obtained, respectively, 
by calculating the one-loop and two-loop
corrections to the two-point Green function which imbeds
$O^{\nu}_nS_{\perp\nu}$.  To obtain $\gamma^{(1)}_n$, 
calculation of 18 two-loop diagrams is required
in the Feynman gauge.
Since the expression for $\gamma^{(1)}_n$ is quite complicated,
we refer the readers to Refs. \cite{Vog,HKK} 
for them.

\begin{figure}[h]
\epsfile{file=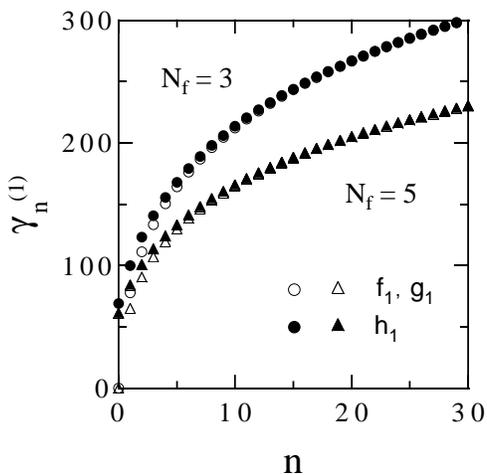,scale=0.7}
\caption[]{The NLO anomalous dimension $\gamma^{h(1)}_n$ in comparison
with
$\gamma^{fg(1)}_n$.  This figure is taken from Ref. \cite{HKK}.}
\end{figure}

In order to get a rough idea about the NLO $Q^2$ dependence
of $h_1$, we plotted
in Fig. 2 $\gamma^{h(1)}_n$ ($\gamma^{(1)}_n$ for $h_1$) in comparison
with $\gamma^{fg(1)}_n$ ($\gamma^{(1)}_n$ for the nonsinglet $f_1$
and $g_1$) for $N_f=3,5$.  One sees from Fig. 2 
$\gamma^{h(1)}_n > \gamma^{fg(1)}_n$ especially at small $n$.
This suggests that the NLO $Q^2$ evolution of
$h_1$ is quite different from that of $f_1$ and $g_1$ in the small
$x$ region.
The relation $\gamma^{h(1)}_n > \gamma^{fg(1)}_n$ is
in parallel with and even more conspicuous than 
the LO anomalous dimensions which read
\bea
\gamma^{h(0)}_n &=& 2C_F\left( 1 + 4 \sum_{j=2}^{n+1}
{1\over j}\right),\nonumber\\
\gamma^{fg(0)}_n &=& 2C_F\left( 1 -{2\over (n+1)(n+2)}
+ 4 \sum_{j=2}^{n+1}{1\over j}\right).
\label{anoLO}
\eea
To illustrate the generic feature of the $Q^2$ evolution,
we have applied the obtained $Q^2$ evolution to a reference
distribution for $g_1$ and $h_1$.
As a reference distribution, we take GRSV $g_1$ distribution\,\cite{GRV2}
and assume $h_1(x,\mu^2)=g_1(x,\mu^2)$ at a low energy input
scale ($\mu^2=0.23$ GeV$^2$ for LO and
$\mu^2=0.34$ GeV$^2$ for NLO evolution) as is suggested by a 
nucleon model\,\cite{JJ,PP}.
We then evolve them to $Q^2=20$ GeV$^2$ and see how much
deviation is produced between them.  The result is shown in Fig. 3.
As is expected from the anomalous dimension,
the drastic difference
in the $Q^2$ evolution between $h_1$ and $g_1$
is observed in the small $x$ region, and this tendency is more
significant for the NLO evolution.\,\cite{BCD,SV2,HKK2} 
(Although $g_1$ for $u$-quark
mixes with the gluon distribution, the same tendency 
in the difference from $h_1$ is observed
for the nonsinglet distribution.)

\hspace{-0.5cm}
\begin{figure}[h]
\epsfile{file=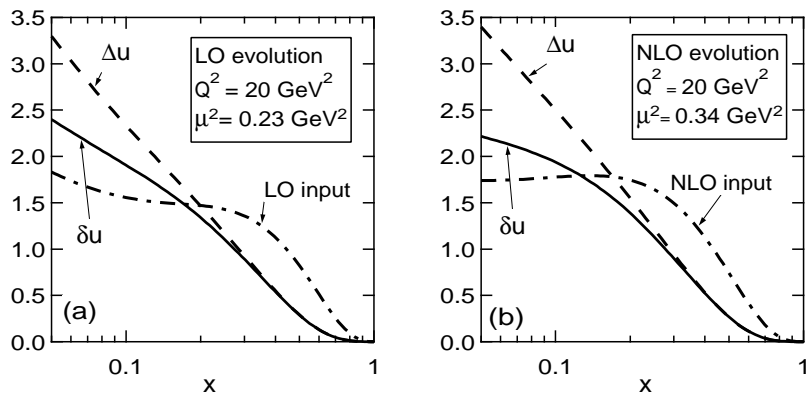,scale=0.57}
\caption[]{(a) The LO $Q^2$ evolution of $h_1$ (denoted by $\delta u$)
and $g_1$ (denoted by $\Delta u$) for the $u$-quark.
(b) The NLO $Q^2$ evolution of $h_1$ and $g_1$ for the $u$-quark.  
This figure is taken from
Ref. \cite{HKK2}}
\end{figure}

As another example, we showed in Fig.4 
the $Q^2$ evolution of the tensor charge (Fig. 4(a))
and the first moment of $h_1$ and the nonsinglet $f_1$ ($g_1$) (Fig. 4(b)).
Although the NLO effect is sizable for the tensor charge,
it is small for the first moment. (For the lattice QCD calculation of
the tensor charge, see Ref. \cite{ADHK}.)

In Ref. \cite{KMSS}, the Regge asymptotics of $h_1$ was studied and 
the small-$x$ behavior was predicted to be $h_1(x)\sim {\rm constant}$
($x\to 0$).
On the other hand, the rightmost singularity of $\gamma^{h(0)}_n$
and $\gamma^{h(1)}_n$ are, respectively, located at $n=-2$ and $n=-1$
in the complex $n$ plane.
Therefore inclusion of the NLO effect in the DGLAP asymptotics
gives consistent behavior at $x\to 0$ as the Regge asymptotics.
This is in contrast to the (nonsinglet) $f_1$ and $g_1$
distributions, whose LO and NLO DGLAP asymptotics 
are the same.

\hspace{-0.5cm}
\begin{figure}[h]
\epsfile{file=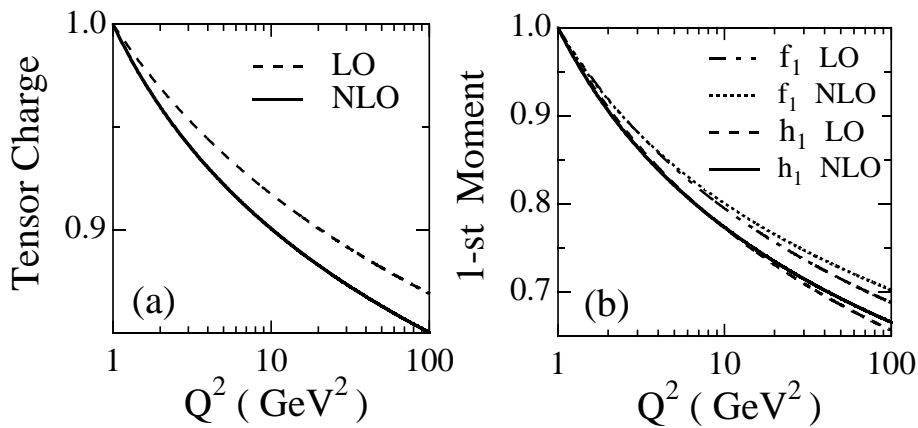,scale=0.65}
\caption[]{(a) The LO and NLO $Q^2$ evolution of the tensor charge
normalized at $Q^2=1$ GeV$^2$.  (b) The LO and NLO $Q^2$ evolution
of the first moment of $h_1$ and the nonsinglet $f_1$ (or $g_1$). 
This figure is taken from Ref. \cite{HKK}.
}
\end{figure}

One of the interesting applications of the 
obtained NLO $Q^2$ evolution of $h_1$
is the preservation of the Soffer's inequality,\cite{Soffer} 
$2|h_1^a(x,Q^2)| \le f_1^a(x,Q^2) +
g_1^a(x,Q^2)$.  
Although the validity of this inequality hinges on schemes beyond
LO\,\cite{GJJ}, the NLO $Q^2$ evolution maintains 
the inequality at
$Q^2> Q_0^2$ if it is satisfied at some (low) scale $Q_0^2$ 
in suitably defined factorization schemes such as 
$\overline{\rm MS}$ and Drell-Yan factorization schemes.\,\cite{BST,Vog}.

As was discussed in Sec. 2, a physical cross section
is a convolution of a parton distribution and a short distance cross section.
(See (\ref{DIS}) and (\ref{DY})) 
For the double transverse spin asymmetry ($A_{TT}$) 
in the Drell-Yan process,
the NLO short distance cross section has been calculated
in Ref. \cite{CKM} in the $\overline{\rm MS}$ scheme.
The analysis on $A_{TT}$ combined with the NLO
tranversity distribution 
predicts modest but not negligible NLO effect.\cite{MSSV}

\section{$Q^2$-evolution of
$h_L(x,Q^2)$ and its $N_c\to\infty$ limit}

In general, higher twist ($\tau \ge 3$) distributions
represent quark-gluon correlation in the nucleon.
Using the QCD equation of motion,
one obtains from (\ref{h1hL}) the following relation\,\cite{BBKT}:
\bea
h_L(x,\mu^2)&=&
2x\int_x^1{dy\over y^2}h_1(y,\mu^2) + \widetilde{h}_L(x,\mu^2),
\label{hL}\\
\widetilde{h}_L(x,\mu^2)&=&{iP^+\over M}
\int_{-\infty}^\infty{dz^-\over 2\pi}e^{-2ixP\cdot z}
\int_0^1udu\int_{-u}^u tdt
\nonumber\\
& &\times
\la PS_\parallel|\bar{\psi}(uz)i\gamma_5\sigma_{\mu\alpha}gG_\nu^{\ \alpha}
(tz)z^\mu z^\nu\psi(-uz)|PS_\parallel\ra,
\label{hLtilde}
\eea
where $z^2=0$, $z^+=0$ and $S_\parallel$ stands for the
longitudinal polarization for the nucleon
($S^\mu=S_\parallel^\mu=p^\mu -{M^2\over 2}n^\mu$).
This equation means that $h_L$ consists of
the twist-2 contribution and  
$\widetilde{h}_L$ which represents quark-gluon correlation
in the nucleon.  We call the latter contribution
``purely twist-3'' contribution.  (Expansion of (\ref{hLtilde})
produces twist-3 local operators.  See (\ref{eq12}) below.)
Equation (\ref{hL}) reminds us of the Wandzura-Wilczek
relation\,\cite{WW} for $g_T$:
\bea
g_T(x,\mu^2)=\int_x^1{dy\over y}g_1(y,\mu^2) + \widetilde{g}_T(x,\mu^2).
\eea
For $e$ and $\widetilde{g}_T$, one can write down relations similar
to (\ref{hLtilde}).

The $Q^2$-evolution of the first and second terms in (\ref{hL})
is described separately.  The evolution of
$\widetilde{h}_L$ is quite complicated.
A detailed analysis of (\ref{hLtilde}) leads to the
following relation for the $n$-th moment
of $\widetilde{h}_L$\,\cite{JJ}:
\bea
& &{\cal M}_n[\widetilde{h}_L(\mu^2)] = \sum_{k=2}^{[(n+1)/2]}
\left( 1 - {2k \over n+2}\right) {1\over 2M}
\la PS_\parallel |R_{nk}(\mu)
|PS_\parallel\ra,
\label{eq11}\\
& &R_{nk}={1\over 2}\left[
\bar{\psi}\sigma^{\lambda\alpha}n_\lambda
i\gamma_5 (in\cdot D)^{k-2}
igG_{\nu\alpha}n^\nu(in\cdot D)^{n-k}\psi - (k\to n-k+2)\right]. \nonumber\\
\label{eq12}
\eea
We note that the number of independent operators
$\{ R_{nk} \}$
($k=2,\cdots,[(n+1)/2]$) increases with $n$.
In the $Q^2$-evolution,
the mixing among
$\{ R_{nk} \}$ occurs and the renormalization is
described by the
anomalous dimension matrix $[\gamma_n(g)]_{kl}$
for $\{ R_{nk} \}$.
If we put the LO anomalous dimension matrix for $\{ R_{nk} \}$ as
$[\gamma_n(g)]_{kl}=(\alpha_s/2\pi)[X_n]_{kl}$
corresponding to (\ref{eq8}),
the solution to the renormalization group equation for
$\{ R_{nk} \}$ takes the following matrix form:
\bea
\la PS_\parallel | R_{nk}(Q^2)|PS_\parallel\ra= \sum_{l=2}^{[(n+1)/2]}
\left[ L^{X_n/\beta_0}\right]_{kl}
\la PS_\parallel | R_{nl}(\mu^2)|PS_\parallel\ra,
\label{eq13}
\eea
where $L\equiv {\alpha_s(Q^2)\over \alpha_s(\mu^2)}$.
$X_n$ for $\widetilde{h}_L$ was derived in Ref. \cite{KT}.
The $Q^2$-evolution for $\widetilde{g}_T$ and 
$e$ is also described by matrix equation similar to
(\ref{eq13}), and the solution was obtained in 
Refs.\,\cite{SV,KYTU} for $g_T$
and in Ref.\,\cite{KN} for $e$.\,\cite{Comment}
As is clear from 
(\ref{eq11}) and (\ref{eq13})
${\cal M}_n[\widetilde{h}_L(Q^2)]$
and ${\cal M}_n[\widetilde{h}_L(\mu^2)]$
are not connected by a simple equation 
as in the case for the twist-2 distribution (see (\ref{eq6})).\cite{BM2}
Although (\ref{eq13}) gives complete prediction
for the $Q^2$ evolution,
it is generally 
difficult to distinguish 
contribution from many operators in the analysis
of experiments.

\begin{figure}[h]
\epsfile{file=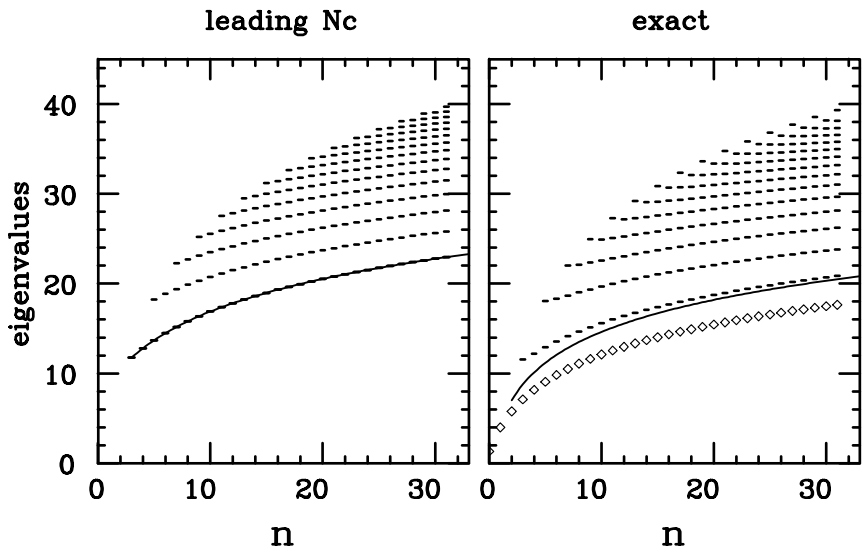,scale=0.60}
\caption[]{(Right) Complete spectrum of the 
eigenvalues of the anomalous dimension
matrix for $\widetilde{h}_L$
obtained in Ref. \cite{KT}. The symbol $\diamond$
denotes the one-loop anomalous dimension for $h_1$.  The solid line
is the anomalous dimension (\ref{largen}) at large $n$.
(Left) Spectrum of the eigenvalues of the anomalous dimension matrix for 
$\widetilde{h}_L$ at large $N_c$.  The solid line denotes the analytic 
solution given in (\ref{eq15}).  
This figure is taken from Ref. \cite{BBKT}.}
\end{figure}

In order to get some feeling on the $Q^2$-evolution
of $\widetilde{h}_L$, we plotted the eigenvalues of $X_n$
in Fig. 5 (right).  For comparison, we also showed in the same figure
the LO anomalous dimension $\gamma_n^{(0)}/2$ for $h_1$.
(Note the differene in convention between (\ref{eq6}) and (\ref{eq13}).)
As is clear from this figure, the $Q^2$ evolution
of $\widetilde{h}_L$ is much faster than that of $h_1$.
(See discussion below.)

In the recent literature\,\cite{ABH,BBKT,KN}, 
it has been shown that at large 
$N_c$ (the number of colors), a great simplification
occurs in the $Q^2$-evolution of the twist-3 distributions.
Recall $X_n$ in (\ref{eq13}) is a function of two Casimir operators
$C_G=N_c$ and $C_F={N_c^2-1 \over 2N_c}$.
If one takes $N_c \to \infty$, i.e. $C_F\to N_c/2$,
(\ref{eq11}) and (\ref{eq13}) is reduced to
\bea
{\cal M}_n[\widetilde{h}_L(Q^2)]
&=& L^{\widetilde{\gamma}^h_n/\beta_0}{\cal M}_n[\widetilde{h}_L(\mu^2)],
\label{eq14}\\
\widetilde{\gamma}^h_n&=&2N_c \left( 
\sum_{j=1}^n{1\over j} - {1\over 4} +{ 3 \over 2(n+1)}
\right).
\label{eq15}
\eea
This evolution equation is just like those for the
twist-2 distributions (see (\ref{eq6})).  In Fig. 5 (left), we showed 
the distribution
of the eigenvalues of $X_n$ 
obtained numerically at $N_c\to \infty$.  The solid line is the
analytic solution in (\ref{eq15}), which shows (\ref{eq15}) 
corresponds to the lowest eigenvalues at $N_c\to\infty$.
Since (\ref{eq14}) was obtained by a mere replacement
$C_F\to N_c/2$ in (\ref{eq13}), the correction to
the result is of $O(1/N_c^2)\sim 10$ \% level, which 
gives enough accuaracy for practical applications.

This large-$N_c$ simplification is a consequence
of the fact that the coefficients of $R_{nk}$ in (\ref{eq11})
constitutes the {\it left} eigenvector of $X_n$ 
corresponding to the eigenvalue $\widetilde{\gamma}^h_n$ in this limit:
\bea
\sum_{k=2}^{[(n+1)/2]}\left( 1 - {2k\over n+2}\right)[X_n]_{kl} =-
\left( 1 - {2l\over n+2}\right)\widetilde{\gamma}^h_n,
\eea
which implies that all the {\it right} eigenvectors of $X_n$
except the one corresponding to $\widetilde{\gamma}^h_n$
are orthogonal to the vector consisting of 
$\left( 1 - {2k\over n+2}\right)$.  This leads to (\ref{eq14}).

This large-$N_c$ simplification of the $Q^2$ evolution was 
proved for the nonsinglet 
$\widetilde{g}_T$ 
in Ref.\,\cite{ABH} and for $\widetilde{h}_L$ and $e$
in Ref.\,\cite{BBKT}.  The corresponding anomalous dimensions 
for $\widetilde{g}_T$ and $e$ are, respectively, 
\bea
\widetilde{\gamma}^g_n&=&2N_c \left( 
\sum_{j=1}^n{1\over j} - {1\over 4} +{ 1 \over 2(n+1)}
\right),\nonumber\\
\widetilde{\gamma}^e_n&=&2N_c \left( 
\sum_{j=1}^n{1\over j} - {1\over 4} -{ 1 \over 2(n+1)}
\right).
\label{largeNc}
\eea

Corresponding to three twist-3 distributions in table 1,
there are three independent twist-3 fragmentation functions.\,\cite{Ji94}
(Their number is doubled to 6 if one includes
final state interactions.  See Ref.\,\cite{Ji94}) 
It has been shown in Ref.\,\cite{AVB} that at large $N_c$
the $Q^2$ evolution of 
all these nonsinglet fragmentation functions is 
also described by a simple evolution equation similar to (\ref{eq14}).
Therefore the simplification of the twist-3 evolution equation
is universal to all twist-3 nonsinglet 
distribution and fragmentation functions.

To illustrate the actual $Q^2$ evolution of $h_L$, 
we have applied (\ref{eq14})
to the bag model calculation of $h_L$.\,\cite{KK} (Fig. 6)  
Fig. 6(a) shows the bag calculation of $h_L$\,\cite{JJ}.
At the bag scale, purely 
twist-3 contribution $\widetilde{h}_L$ is comparable to
the twist-2 contribution.  After the $Q^2$ evolution to
$Q^2= 10$ GeV$^2$, $h_L$ is dominated by the twist-2 contribution.
This is a consequence of the large anomalous dimension
(\ref{eq15}) compared with the LO anomalous dimension
of $h_1$ ($\diamond$ in the right figure of Fig. 5).
A similar calculation was done for $g_T$ in Ref.\,\cite{St}.

\begin{figure}[h]
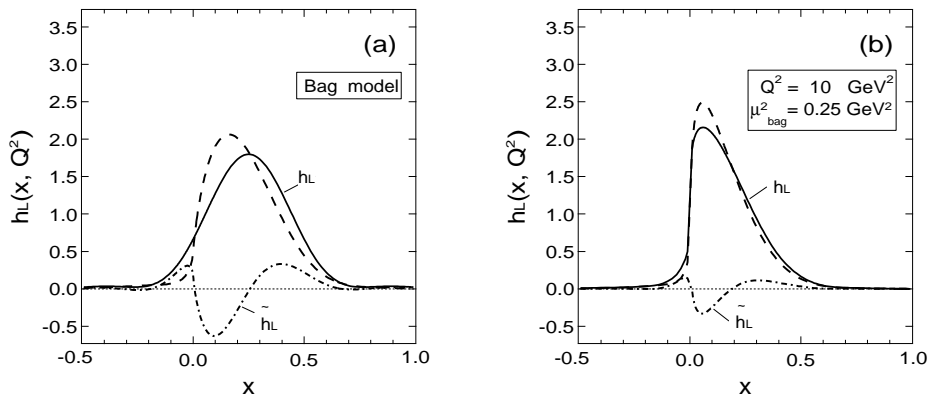

\epsfile{file=fig6.eps,scale=0.65}
\caption[]{(a) Bag model prediction for $h_L$. The dashed line 
represents the twist-2 contribution to $h_L$.
(b) Bag model prediction for $h_L$ evolved to $Q^2=10$ GeV$^2$
assuming the bag scale is $\mu^2=0.25$ GeV$^2$.  These figures are taken
from Ref. \cite{KK}}
\end{figure}

Another simplification of the twist-3 evolution occurs
at $n\to\infty$.\cite{ABH,BBKT}  In this limit, all the twist-3
distributions obey a simple DGLAP equation (\ref{eq14}) with 
a common anomalous dimension which is slightly shifted
from (\ref{eq15}) and (\ref{largeNc}):
\bea
\gamma_n = 4C_F\left(\sum_{j=1}^n{1\over j} - {3\over 4} \right) + N_c.
\label{largen}
\eea
This evolution equation satisfies the complete evolution equation 
to the $O({\rm ln}(n)/n)$ accuracy\,\cite{ABH}.
In the right figure of Fig. 5, (\ref{largen}) is shown by the solid line.
One sees that it is close to the lowest eigenvalues except for small $n$. 
Combined with this $n\to \infty$ result, the 
large-$N_c$ evolution equation
in (\ref{eq14}) with (\ref{eq15}) and (\ref{largeNc}) for each distribution
is valid to $O((1/N_c^2){\rm ln}(n)/n)$ accuracy.

\section{Summary}

In this talk, I discussed the recent progress 
in perturbative QCD, especially the
$Q^2$ evolution of
the chiral-odd spin-dependent parton distributions $h_1(x,Q^2)$
and $h_L(x,Q^2)$.

The NLO $Q^2$ evolution for
the transversity
distribution $h_1(x,Q^2)$ was completed in the $\overline{\rm MS}$
scheme.  
This means the $Q^2$ evolution of all the
twist-2 distributions has been understood in the NLO level.
The resulting $Q^2$ evolution of $h_1(x,Q^2)$ turned out to cause
quite different behavior from the helicity distribution $g_1(x,Q^2)$
in the small $x$ region if $h_1$ is assumed to be equal to
$g_1$ at a
low energy scale.

The LO $Q^2$ evolution for the twist-3 distribution $h_L(x,Q^2)$
(and $e(x,Q^2)$)
was completed.  
Although their $Q^2$ evolution is quite complicated
due to the mixing among increasing number of quark-gluon-quark operators,
it obeys a simple DGLAP equation similar to the twist-2 distibution
in the $N_c\to \infty$ limit,
as was the case for the $Q^2$ evolution of the nonsinglet
$g_T(x,Q^2)$ distribution.  The same 
simplification 
at $N_c\to \infty$ was also proved for the
twist-3 fragmentation functions.  Therefore this large-$N_c$ simplification
was proved to be universal for the twist-3 distribution and fragmentation
functions.

By these studies, we now have at hand the necessary tools
given by perturbative QCD 
for the analysis of the whole parton distributions.
From the point of view of the ``spin distributions'' in the nucleon,
one is more interested in the nonperturbative $x$ dependence
of those parton distributions.  Armed with the development 
in the perturbative $Q^2$ dependence and the 
nonperturbative QCD techniques, 
we will be forced to challenge this issue when the new spin colliders
start producing data on the new spin distributions. 

\section*{Acknowledgements}

I would like to thank I.I. Balitsky, V.M. Braun, 
A. Hayashigaki, Y. Kanazawa, N. Nishiyama and K. Tanaka for the
collaboration on which this talk is based.
I'm also greatful to J. Kodaira for inviting me to such an
exciting conference.

\end{document}